\def\ref{\par\noindent\hang}
\def\AaA{{\em Astr.~Astrophys.}}

\def\PhD{{\em PhD thesis}}

\def\etal{{et al.\thinspace}}

\def\ie{{\em i.e.\ }}
\def\spose#1{\hbox to 0pt{#1\hss}}
\def\approxlt{\mathrel{\spose{\lower 3pt\hbox{$\sim$}}
	\raise 2.0pt\hbox{$<$}}}
\def\approxgt{\mathrel{\spose{\lower 3pt\hbox{$\sim$}}
	\raise 2.0pt\hbox{$>$}}}

\def\multleft#1{\hbox to size{\vbox {\halign {\lft{##}\cr #1}}\hfill}\par}
\def\multright#1{\hbox to size{\vbox {\halign {\rt{##}\cr #1}}\hfill}\par}

\def\today{\ifcase\month\or January\or February\or March\or April\or May\or
      June\or July\or August\or September\or October\or November\or December\fi
      \space\number\day, \number\year}
\def\$<${\thinspace}
\def\s{\hbox{\phantom{5}}}	

\def\boxit#1{\vbox{\hrule\hbox{\vrule\kern3pt\vbox{\kern3pt
          #1 \kern3pt}\kern3pt\vrule}\hrule}}

\def\cm{{\rm\thinspace cm}}

\def\erg{{\rm\thinspace erg}}

\def\g{{\rm\thinspace g}}

\def\keV{{\rm\thinspace keV}}
\def\km{{\rm\thinspace km}}

\def\Mpc{{\rm\thinspace Mpc}}

\def\s{{\rm\thinspace s}}

\def\sr{{\rm\thinspace sr}}

\def\cmpssq{\hbox{$\cm\s^{-2}\,$}}

\def\pcmcu{\hbox{$\cm^{-3}\,$}}

\def\ergpcmps{\hbox{$\erg\cm^{-3}\s^{-1}\,$}}

\def\ergps{\hbox{$\erg\s^{-1}\,$}}
\def\ergpspmp{\hbox{$\erg\s^{-1}\Mpc^{-3}\,$}}
\def\gpcm{\hbox{$\g\cm^{-3}\,$}}

\def\gps{\hbox{$\g\s^{-1}\,$}}

\def\kmps{\hbox{$\km\s^{-1}\,$}}

\def\pcmsq{\hbox{$\cm^{-2}\,$}}

\def\pmpccu{\hbox{$\Mpc^{-3}\,$}}
\def\ps{\hbox{$\s^{-1}\,$}}

\def\psr{\hbox{$\sr^{-1}\,$}}

\def\kmpspMpc{\hbox{$\kmps\Mpc^{-1}$}}

\documentstyle[psfig]{mn}

\title[The X-ray Background]
{Ion-supported tori: a thermal bremsstrahlung model for the X-ray Background}

\author[T. Di Matteo and A.~C.~Fabian]
{T. Di Matteo and A.~C.~Fabian\\
Institute of Astronomy, Madingley Road, Cambridge CB3 0HA}

\date{}

\begin{document}

\maketitle

\begin{abstract}
We discuss the possibility that a significant contribution of the hard
X-ray Background is the integrated emission from a population of
galaxies undergoing advection-dominated accretion in their
nuclei. Owing to poor coupling between ions and electrons and to 
efficient radiative cooling of the electrons, the accreting
plasma is two-temperature, with the ions being generally much hotter than
the electrons and forming an ion-supported torus.  We show that the electron temperature then
saturates at $\approx 100 \keV$ independent of
model parameters. At this temperature the hard X-ray emission is
dominated by bremsstrahlung radiation. We find that this physical
model gives an excellent fit to the spectrum of the XRB in the 3-60 \keV ~range, provided that there is
 some
evolution associated with the spectral emissivity which must peak at a redshift $\sim 2$. ~We estimate
that such galaxies contribute only to a small 
fraction of the local X-ray  volume emissivity. The model implies a higher mean black hole mass than is obtained from the evolution of quasars alone.
\end{abstract}

\begin{keywords} galaxies: active -- galaxies: nuclei; accretion, accretion disks -- X-rays: general -- galaxies
\end{keywords}

\section{Introduction}
The puzzle of the origin of the X-ray-background (XRB)  remains unsolved after over 
30 years of study. Thermal bremsstrahlung with a temperature of about
40 \keV ~fits its spectrum between 3 and 60 \keV ~very well (Marshall \etal
 1980). Despite this agreement, observations of the cosmic microwave background (CMB) exclude
the possibility that hot intergalactic gas is a major contributor to
the XRB. The excellent fit of the CMB to the blackbody function
 strictly limits the amount of hot diffuse gas between the current
epoch and the last scattering surface at redshift $z\approx 1000$; such gas  would
distort the XRB spectrum through the Compton scattering. The upper
limit to any such  distortion gives an upper limit to the contribution of
the hot intergalactic gas to the XRB as $\approxlt 3$ ~per cent (Mather \etal
1990).  
\par It is therefore likely that the XRB is due to a
contribution of different types of discrete sources. Active galactic
nuclei (AGN) are likely candidates (see Fabian and Barcons 1992 and references therein), especially
since Seyferts and Quasars (QSO) provide a large fraction of the soft
XRB below $\sim$ 2 \keV ~(Shanks \etal 1991; Hasinger \etal 1993). One of the major problems
with an AGN origin of the XRB has been the apparent discrepancy of the
XRB spectrum with the spectrum of resolved AGN. In the $2-10 \keV$ ~band
AGN spectra are too steep and subtraction of their contribution worsens the situation (this is the spectral paradox of Boldt
1987). Moreover detailed studies of the QSO X-ray luminosity function
(Boyle \etal 1994) and the source number count distribution have shown
that QSOs are unlikely to form more than 50 per cent of the XRB, even at 1 \keV.
\par
Models based on the unified Seyfert scheme, in which a large fraction of the emission from AGN is absorbed by obscuring matter (Madau \etal 1994; Celotti \etal 1995; Comastri \etal 1995)  seem to work. In these models though, it is difficult to obtain a smooth spectrum in the 1-10 \keV ~band (Matt and Fabian 1993). Some spectral features are expected due to the matter in the vicinity producing an iron edge and emission line in the intrinsic spectrum. The requirement of a smooth spectrum for the XRB has been recently enphasized by ASCA observations which show no such spectral features (Gendreau \etal 1995a, 1995b), implying that a new faint population with a very hard, smooth X-ray spectrum is required.
\par
A promising new population of sources with harder mean X-ray spectra
discovered  in deep ROSAT images (Hasinger \etal 1993; Vikhlinin \etal 
1995; McHardy \etal 1995; Boyle \etal 1995; Almaini \etal 1996) is emerging as a significant 
possibility for the missing hard component of the XRB. At present this
population can account for only about 10 per cent of the XRB, but the source
number counts are still climbing at the lowest detected fluxes. Recent deep ROSAT studies are
beginning to resolve some of the population into  narrow emission-line
galaxies (Boyle \etal 1995; Griffiths \etal 1996) with remarkably high
X-ray luminosities, typically two orders of magnitude more than that of 
 galaxies observed locally (Fabbiano 1989) despite comparable
optical luminosities. Although it now seems clear that faint galaxies
are emerging as a significant new X-ray population, questions on
the origin and nature of their activity still remain unsolved.
\par
The spectrum of the XRB, with its $\sim
30 \keV$ rollover, requires some spectral uniformity in its
constituent sources. Since it resembles  thermal bremsstrahlung
in the 3-60 \keV ~range, a mechanism which produces such radiation in
galaxy nuclei would be particularly appealing. A way of
standardising the temperature is then  required. A possibility
has emerged in relation to recent discussion of energy advection solutions
(Shapiro, Lightman \& Eardley 1976; Begelman 1978; Rees \etal 1982; Abramowicz \etal 1995; Narayan 1996; Chakrabarti 1996) for accretion disks by Narayan
\& Yi (1995a,b). Beginning with the work of
Shapiro, Lightman \& Eardley (1976) and Rees \etal (1982), investigations of black hole
accretion disks at low $\dot{M} (\le 0.01\dot{M}_{\rm Edd})$, have focussed on
a class of optically thin solutions where the gas is significantly
hotter than in the local Shakura-Sunyaev (1973) thin disk solution. Because of the poor radiative efficiency of the accreting gas, in
the advection-dominated solution most of the accretion energy is
stored within the gas and advected radially inward.  The accreting
plasma in this solution is two-temperature; since the ions are
much  hotter than the electrons
they maintain a thick torus (which is supported by the ion pressure). This requires the gas density to be sufficiently low that ion-electron coupling via Coulomb collisions becomes weak and cooling via synchrotron and bremsstrahlung radiation is not very important. In addition, the gas is optically thin for Compton cooling to be modest. For these reasons, the ion-supported torus exists only at low mass accretion rates (Rees \etal 1982). 

 A major attraction of this class of solution is
that the plasma attains and maintains an electron temperature $kT_{\rm e} \approx 100 \keV$
and, since the gas is optically thin, much of the X-ray cooling occurs
through thermal bremsstrahlung. X-ray spectra from such sources, when
integrated over redshift, should plausibly resemble the XRB.
\par
In this paper we explore a model in which a population of
galactic nuclei undergoes advection-dominated accretion, perhaps as
their quasar activity diminishes, and so produces the XRB
spectrum. Although we centre our model on  advection-dominated
disks (ADD), any situation in which an optically thin, two-temperature
magnetized gas occurs would suffice.
\par
In section 2 of this paper we review the physical conditions in ADD, as
discussed in Narayan \& Yi (1995b). We determine the two plasma temperatures
from balancing the heating and cooling, and deduce the process
that dominates the X-ray emission. In section 3, we derive the
spectrum of the XRB from our model. We then discuss our results and
their implications in the last section.
\section{X-ray emission from the advection-dominated disk}
Since most of the accretion energy in advection-dominated flows is stored within the gas and advected radially inward we expect the temperature to be very high and much of the radiation to come out in hard X-rays.
\subsection{Physical properties of advection dominated disks}
Narayan \& Yi (1994,1995a) have investigated the general properties of advection-dominated flows and derived self-similar and height-integrated solutions for the continuity, energy and momentum equations. We review here those equations from their work (and those of Stepney \& Guilbert 1983 and Zdziarski 1988) which are relevant to the present discussion. The solutions give us quantitative estimates for the angular velocity, radial velocities and  density of the flow as a function of radius $R,$ total mass $M$, accretion mass $\dot{M}$,  and other parameters, such as the ratio of specific heats $\gamma$, the viscosity parameter $\alpha$ and the fraction of viscously-dissipated energy which is advected, $f$ (from 0 to 1); a fraction ($1-f$) of the energy is radiated. 
\par
In order to obtain the temperature of the flow we need an equation of state and the equation for the  balance of heating and cooling. We allow the electron temperature $T_{\rm e}$, and the ions temperature $T_{\rm i}$, to be different and we take the gas pressure to be given by 

\begin{equation}
p_g=\beta \rho c_s^2= \frac{\rho kT_{\rm i}}{\mu_{\rm i}m_u}+\frac{\rho kT_{\rm e}}{\mu_{\rm i}m_u}, 
\end{equation}
where $\rho$ is the density of the gas, $c_s$ the isothermal sound speed, $\mu_{\rm i}$ and $\mu_{\rm e}$ the effective molecular weights of the ions and electrons respectively, $\beta$ the ratio of gas pressure to total pressure where

\begin{eqnarray}
&p_{\rm tot} =& p_{\rm g} + p_{\rm m}, \nonumber \\
&p_{\rm g}=&\beta p_{\rm tot}, \nonumber \\
&p_{\rm m}=&(1-\beta)p_{\rm tot}=\frac{B^2}{8\pi},
\end{eqnarray} 
 $p_{\rm m}$ is the magnetic pressure and $B$ the  magnetic
field. We assume the gas pressure is in equipartition with magnetic pressure and  do not include radiation pressure.   By scaling the mass in units of the solar mass, $M=mM_{\odot}$,
the accretion rate in Eddington units, $\dot{M}
=\dot{m}\dot{M}_{\rm Edd}$, where $\dot{M}_{\rm Edd}=L_{\rm Edd}/\eta c^2 =
1.39\times 10^{18} m \gps$, ($\eta$ is the standard accretion efficiency factor), and radii in units of Schwarzschild
radius, $R=rR_{S}, \thinspace R_{S} = 2.95\times 10^{5}m \cm$ we
have the following relations for  $\rho, c_s^2, \tau_{es}$, (Narayan 1995b)

\begin{eqnarray}
&c_{\rm s}^2 =&\phi(f,\beta)r^{-1} \cmpssq, \nonumber \\
&\rho= &\psi(f,\beta) \alpha^{-1}m^{-1}\dot{m}r^{-3/2}\gpcm, n_{\rm e} = \rho/\mu_{\rm e}m_u\pcmcu,\nonumber \\
&\tau_{\rm es} =& n_{\rm e} \sigma_T R =\varphi(f,\beta) \alpha^{-1} \dot{m}r^{-1/2}
\end{eqnarray}
where $\phi, \psi,\varphi$ express the dependences on $\beta$ and $f$ according to the Narayan \& Yi solutions (1995b). $\tau_{es}$ is the scattering optical depth, $\sigma_T$ is the Thomson cross section, $n_{\rm e}$ the number density of electrons and $\alpha$ is the standard Shakura-Sunyaev viscosity parameter.    
\subsection{Heating and cooling} 
We determine the ion and electron temperatures in the accreting plasma by taking into account the detailed balance of heating and cooling. Because almost all of the energy goes into internal energy, the gas becomes much hotter than in the thin disk solution. In particular, if the heating is adiabatic, then $T_{\rm i}\propto n^{2/3}$ while $T_{\rm e}\propto n^{1/3}$ so the ions are preferentially heated. The viscous dissipation of energy acts primarily on the ions, which transfer little of their energy to the electrons. We assume that the transfer from ions to electrons occurs only through Coulomb coupling. The cooling of the plasma is via electrons and occurs through a variety of channels.
\subsubsection{Coulomb heating of electrons by ions}
The Coulomb rate of transfer of energy in the case of hot ions heating cooler electrons is given by (Stepney \& Guilbert 1983)

\begin{eqnarray}
Q^{+} = \frac{3}{2}\,\frac{m_{\rm e}}{m_{\rm p}}n_{\rm e}n_{\rm i}\sigma_{\rm T}c\frac{(kT_{\rm i}-kT_{\rm e})}{K_2(1/\theta_{\rm e})K_2(1/\theta_{\rm i})}\ln{\Lambda} \nonumber \\
\times \left[\frac{2{(\theta_{\rm e}+\theta_{\rm i})}^2 + 1}{\theta_{\rm e} + \theta_{\rm i}}K_1\left(\frac{\theta_{\rm e} + \theta_{\rm i}}{\theta_{\rm e}\theta_{\rm i}}\right)+2 K_0\left(\frac{\theta_{\rm e} + \theta_{\rm i}}{\theta_{\rm e}\theta_{\rm i}}\right)\right] \nonumber \\
\ergpcmps,
\end{eqnarray}
where ${\rm ln}\Lambda \sim 20$, is the Coulomb logarithm, the $K$'s are the modified Bessel functions, and $\theta_{\rm e},\theta_{\rm i}$ are the dimensionless electron and ion temperatures defined by
\begin{equation}
\theta_{\rm e} = \frac{kT_{\rm e}}{m_{\rm e}c^2},\hspace{.3in} \theta_{\rm i}= \frac{kT_{\rm i}}{m_{\rm i}c^2}.\end{equation}

Electrons are mainly cooled by bremsstrahlung, cyclo-synchrotron  and Compton cooling via interaction with soft photons.
\subsubsection{Bremsstrahlung cooling}
The total bremsstrahlung cooling rate per unit volume is the sum of electron-ion and electron-electron bremsstrahlung
\begin{equation}
Q_{\rm br}^{-} = Q_{\rm eiB}^{-} + Q_{\rm eeB}^{-}.
\end{equation}
According to Stepney \& Guilbert (1983) and taking into account the corrections to the numerical constants  by Narayan \& Yi (1995b) we adopt
\begin{equation}
Q_{\rm eiB}^- = 1.25n_{\rm e}^2\sigma_Tc\alpha_fm_{\rm e}c^2F_{\rm ieB}(\theta_{\rm e})\hspace{.2in}\ergpcmps ,
\end{equation}
$\alpha_f$ is the fine structure constant and the function $F_{\rm ei}(\theta_{\rm e})$  for both $\theta_{\rm e}<1$ and $\theta_{\rm e}>1$ is given in Narayan \& Yi (1995b).
For electron-electron bremsstrahlung we have the following expressions (Gould 1980; Narayan \& Yi 1995b)

\begin{eqnarray}
\theta_{\rm e}<1 \nonumber\\
Q_{\rm eeB}^{-} &= &n_{\rm e}^2r_{\rm e}^2m_{\rm e}c^3\alpha_{\rm f}\frac{20}{9\pi^{1/2}}(44-3\pi^2)\theta^{3/2} \nonumber \\
& &\times(1 + 1.1\theta_{\rm e} + \theta_{\rm e}^2-1.25\theta_{\rm e}^{5/2})
\hspace{0.05in} \ergpcmps,\\
\theta_{\rm e}>1 \nonumber \\
Q_{\rm eeB}^{-}&=&n_{\rm e}^2r_{\rm e}^2m_{\rm e}c^3 \alpha_{\rm f} 24\theta_{\rm e}({\rm ln}2\eta\theta_{\rm e}+1.28)
\nonumber\\ & & \hspace{1.6in}\ergpcmps,
\end{eqnarray}
where $r_{\rm e} = e^2/m_{\rm e}c^2$ is the classical electron radius and $\eta= {\rm {\rm exp}}(-\gamma_{\rm E})$ and $\gamma_{\rm E}$ is Euler's constant $\sim0.5772$.

\subsubsection{Cyclo-synchrotron cooling}
Due to the assumption of an equipartition magnetic field in the plasma, synchrotron emission becomes an important cooling process. The cyclo-synchroton harmonics are self absorbed up to an energy, $x_t$, at which self-absorption ceases to establish a local Planck spectrum, (Zdziarski 1988)
\begin{equation}
x_{\rm t} = 6x_{\rm c}\theta^2{\rm ln}^3\frac{C}{{\rm ln}\frac{C}{{\rm ln}\frac{C}{{\rm ln}\frac{C}\ldots}}},
\end{equation}
where $x_{\rm t}= h\nu_{\rm t}/m_{\rm e}c^2$,
\begin{equation}
C=\frac{1}{2\theta_{\rm e}}\left[\frac{2\pi^{1/2}\tau_{\rm es}{\rm exp}(1/\theta_{\rm e})}{3\theta_{\rm e}^{1/2}\alpha_{\rm f}x_{\rm c}}\right]^{1/3},
\end{equation}
and $x_{\rm c}$ is the dimensionless cyclo-synchroton frequency for magnetic field in equipartition.

\begin{figure}
\centerline{\psfig{figure=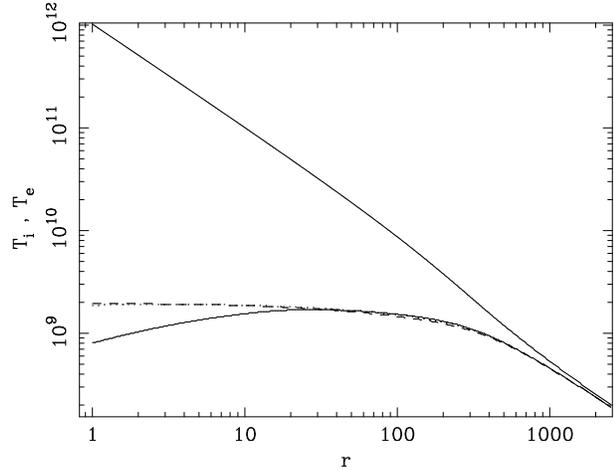,width=0.5\textwidth,angle=270}}
\caption{The plot shows the variations of ion temperature $T_{\rm i}$ and electron temperature $T_{\rm e}$ for three different values of $\dot{m}$ below the critical accretion rate ($\dot{m}_{\rm crit}\approxlt 10^{-1.5}$ $-$see also Narayan \&Yi 1995b). The three different $\dot{m}$ ~($10^{-1.7}$  $-$solid line, $10^{-2.2}$  $-$dotted line, $10^{-2.7}$  $-$dashed line)  models give very similar temperature solutions, in particular the temperature solutions become virtually identical as $\dot{m}$ decreases from $\dot{m}= 10^{-2.2}$ to $10^{-2.7}$. The accreting plasma is single temperature and virial at large radii $r\approxgt10^{3}$. $T_{\rm i} $ remains nearly virial, but $T_{\rm e}$ saturates because of poor coupling between the ions and the electrons and due to a variety of efficient cooling mechanisms for the electrons.} 
\end{figure}

To estimate the cooling per unit volume resulting from the
cyclo-synchroton emission we adopt the expression derived by Narayan
\& Yi (1995). This expression has been obtained by equating the
cyclo-synchroton emission to the Rayleigh-Jeans blackbody emission and
by assuming that at each frequency $\nu$ the observer sees a blackbody
source with a radius determined by the condition $\nu= \nu_{\rm t}(R)$ so
that
\begin{equation}
Q_{\rm synch}^{-} \approx \frac{2\pi}{3}m_{\rm e}\theta_{\rm e}\frac{\nu_{\rm t}^3}{R}\hspace{0.4in}\ergpcmps.
\end{equation}
 
\begin{figure*}
\centerline{
\hbox{
\psfig{figure=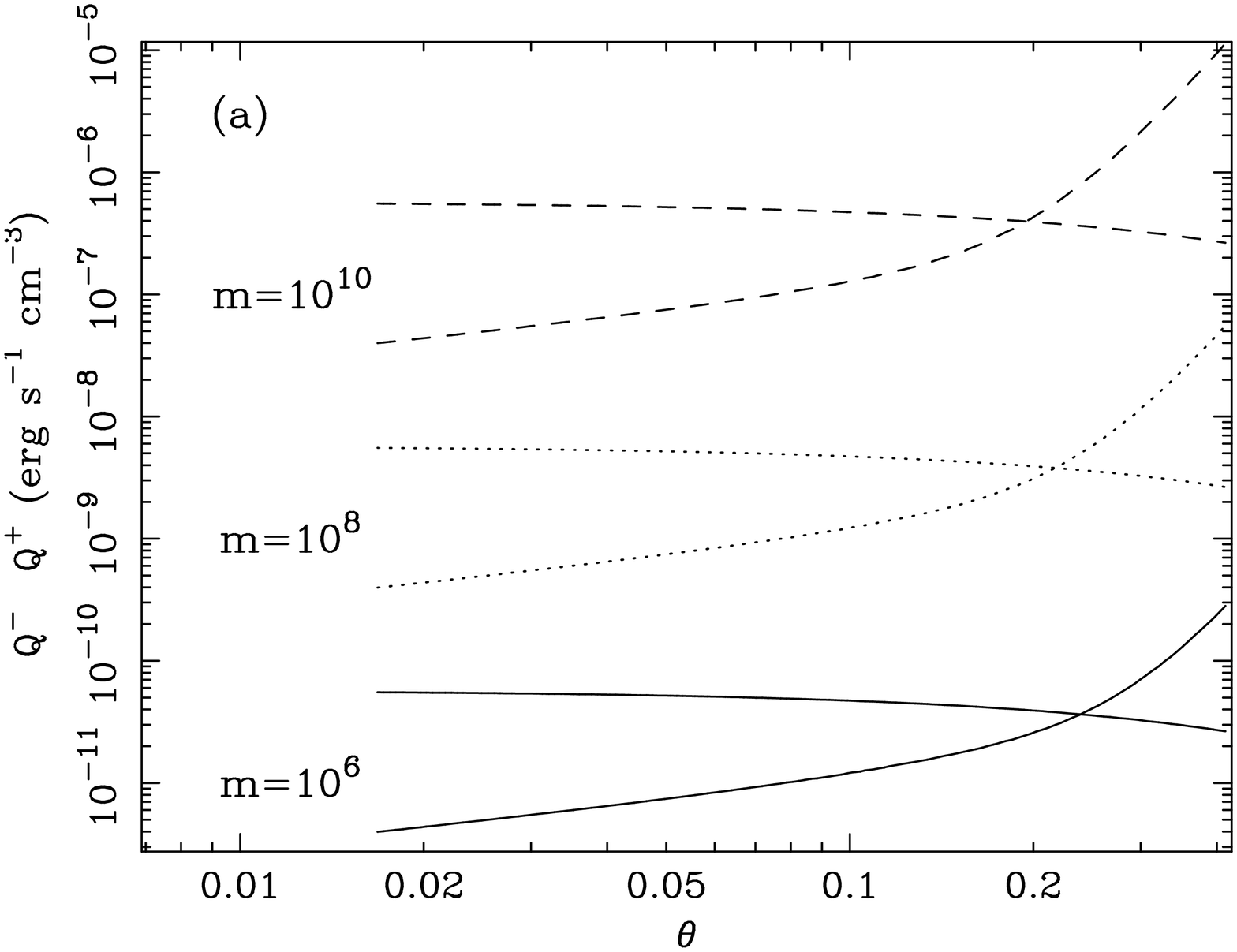,width=0.36\textwidth,angle=270}
\psfig{figure=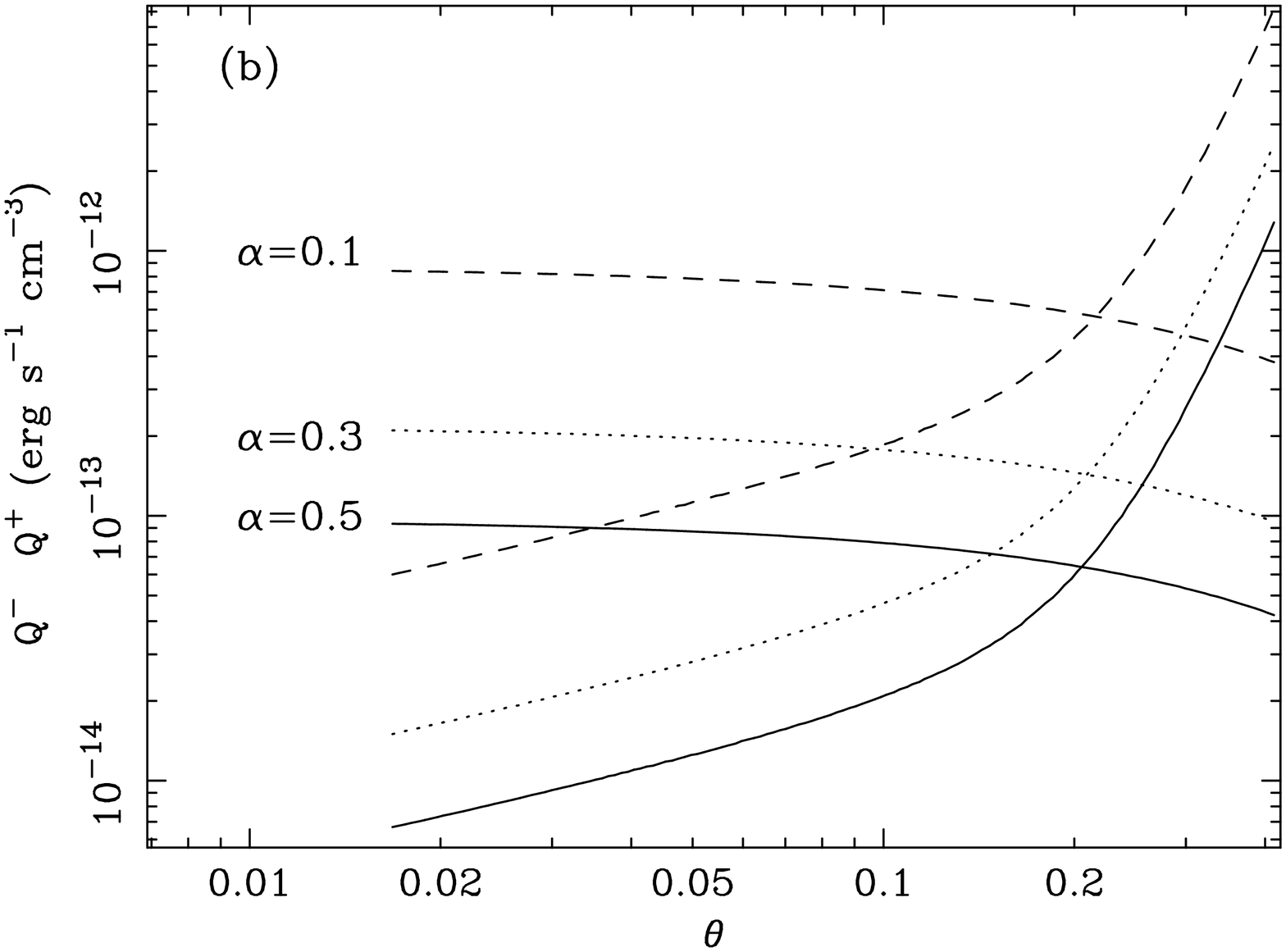,width=0.36\textwidth,angle=270}
\newline
\psfig{figure=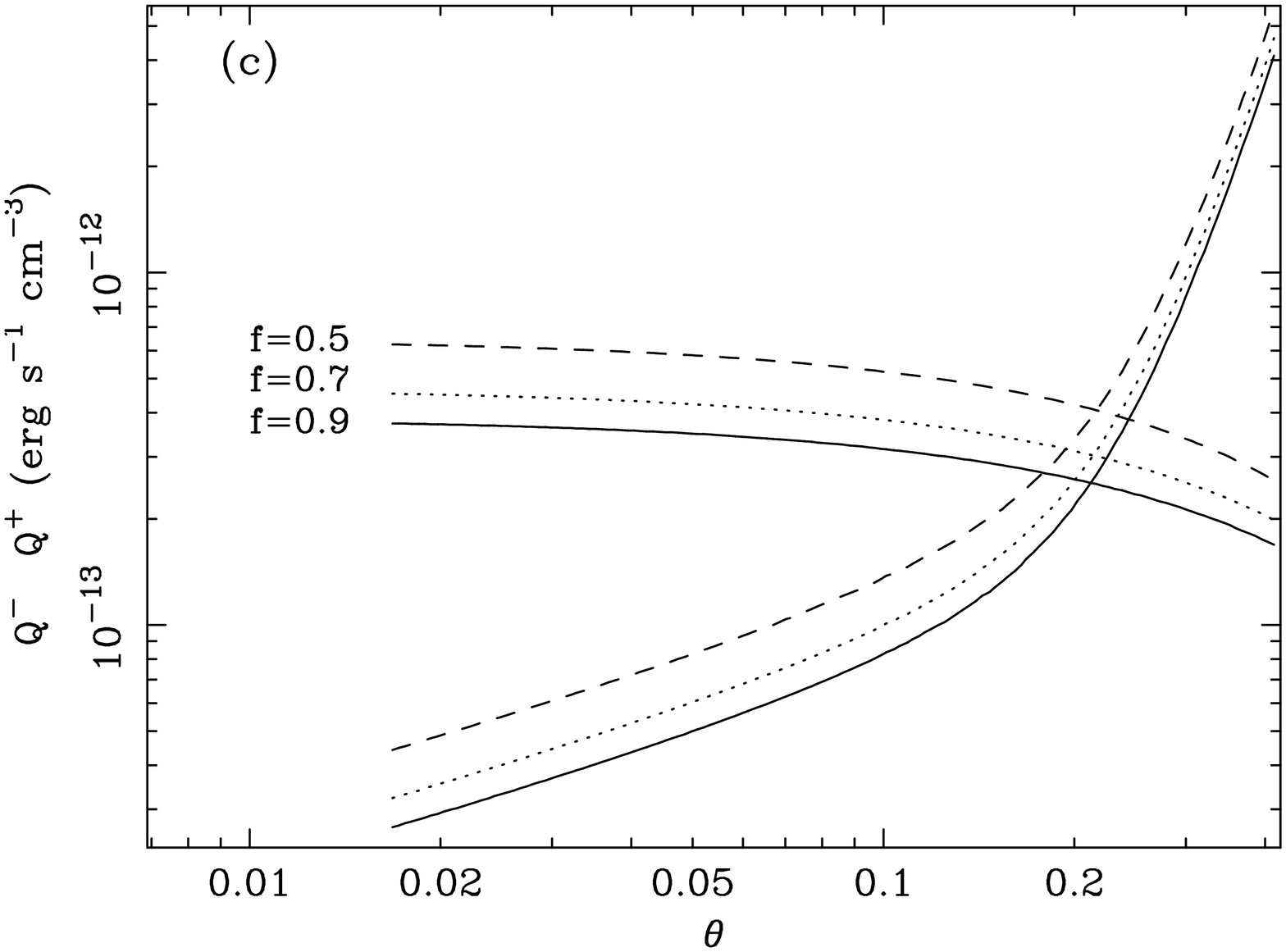,width=0.36\textwidth,angle=270}
}}
\caption{The electron temperature solution is virtually independent of the model parameters. The upper curves in each plot represent the heating function. The lower curves the cooling function. In each of the three plots the intersection point of the respective curves determines the temperature  solution for different values of the parameters. In (a) we vary the black-hole mass $m$, (b) the viscosity coefficient $\alpha$, and (c) the fractional value of $f$, the advection parameter.}
\end{figure*}

\subsubsection{Comptonization of cyclo-synchroton radiation}
Comptonization of soft photons becomes an important cooling mechanism in the inner regions of disk. Comptonization can be parametrized by means of the energy enhancement factor $\eta$ (Dermer \etal 1991, Narayan \& Yi 1995b).
\begin{eqnarray}
\eta = 1 + \frac{P(A - 1)}{1-PA}\left[1-\left(\frac{x}{3\theta_{\rm e}}\right)^{-1-\frac{{\rm ln}{P}}{{\rm ln}{A}}}\right], \nonumber \\ \equiv 1 + \eta_1 -\eta_2\left(\frac{x}{\theta_{\rm e}}\right)^{\eta_3},
\end{eqnarray}
where
\begin{eqnarray}
x&=&hv/m_{\rm e}c^2, \nonumber\\
P&=&1-{\rm exp}(-\tau_{\rm es}), \nonumber\\
 A& =& 1 + 4\theta_{\rm e} + 16\theta_{\rm e}^2.
\end{eqnarray}
The factor $P$ is the probability that an escaping photon is scattered, while $
A$ is the mean amplification factor in the energy of a scattered photon when the the scattering electrons have a Maxwellian velocity distribution.
\par
Most of the cyclo-synchroton radiation is emitted near the self-absorption energy $x_{\rm t}$ given by equation (10). The cooling rate due to comptonization of this radiation is given by
\begin{equation}
Q_{\rm Comp}^{-}= Q_{\rm synch}^{-}\left[\eta_1-\eta_2\left(\frac{x_{\rm t}}{\theta_{\rm e}}\right)^{\eta_3}\right].
\end{equation}
\subsection{The plasma temperatures: thermal balance of  electrons}
 The net volume cooling rate for the accreting gas equals the sum of the three cooling rates given above, \ie 
\begin{equation}
Q^-=  Q_{\rm synch}^{-}+ Q_{\rm Comp}^{-} +Q_{\rm br}^{-}.
\end{equation} 
 The energy balance of the electrons requires the net heating rate to be equal to the net cooling rate, this gives
\begin{equation}
Q^{+}= Q^- .
\end{equation}

This relation, together with equation (1) is  solved to obtain $T_{\rm i}$ and $T_{\rm e}$,
for a given $m,\dot{m},f,\alpha$, and $\beta$. In order to characterize the  solution for advection-dominated AGN, we choose $m = 10^7, \alpha =0.3, f=0.9$ for  and $\beta = 0.5$. 
A plot of the temperatures $T_{\rm i}$ and $T_{\rm e}$  as a function of radius and for three different values of $\dot{m}$ is given in Fig.~1.  The ions achieve nearly  virial temperature, $T_{\rm i} = 10^{12}r^{-1}$ K, and the electrons are cooler, $T_{\rm e} \approx 10^{9}$ K for any given $\dot{m}$ below the critical value, $\dot{m}\approxlt 0.1\alpha^2$ ~(Rees \etal 1982; Narayan \& Yi 1995) for which advection-dominated solution would not exist, and for $r \approxlt 10^3$. 
\par
The stability of the solution for the electron temperature $T_{\rm e}
\approx 10^{9}$ K for $r\approxlt 10^3$ is a particularly strong property of the model.  In
Fig. 2 we plot the total cooling rate $Q^-$ and the total heating rate
$Q^{+} $as function of temperature and we let the different
parameters $m,\dot{m},f,\alpha$, and $\beta$ vary in the respective
ranges of interest. These plots show that the temperature solution is
essentially independent of model parameters as the energy balance
solution, represented by the intersection point of the heating and
cooling functions, always occurs at $\theta_{\rm e} = 0.20\pm0.03$ in the
different plots. This is related to the fact that the slope of the cooling function is steeply increased by the  $Q^-_{\rm synch}$ contribution just above  $\theta_{\rm e}=0.2$ ($Q^-_{\rm synch}$ is the strongest function of $\theta_{\rm e}$; Eqns. (10)-(12)).
\begin{figure}
\centerline{\psfig{figure=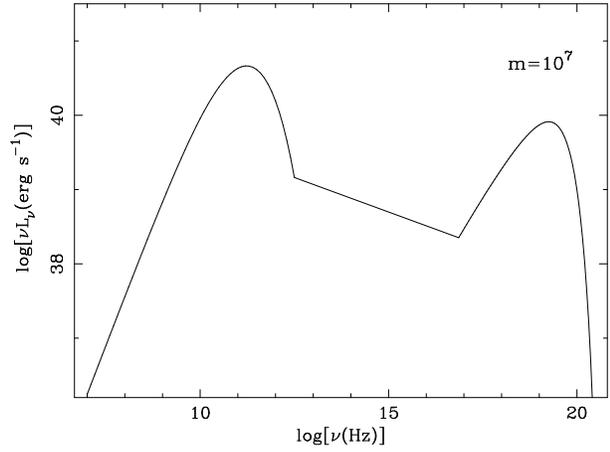,width=0.5\textwidth,angle=270}}
\caption{Emission produced by accretion onto a $10^7M_{\odot}$ black-hole. Here we show the two major peaks in the emission corresponding to the two most efficient cooling mechanisms for the electrons, i.e. synchrotron and bremsstrahlung. The Compton emission is roughly indicated by the line between the two peaks.} 
\end{figure}

We calculate the spectrum for the hot-ion sources  including cyclo-synchrotron emission and bremsstrahlung (Fig. 3) given the model parameters described above.    
We do not include the detailed spectrum due to Compton scattering of synchrotron photons as this is complex and not relevant for the present discussion. The bulk of the emission occurs in the high radio band close to the self-absorption frequency $ \nu_{\rm t}\approx 6\times10^{11}$ Hz. The position of the bremsstrahlung
peak is determined primarily by $T_{\rm e}$.
At the  electron temperature $\theta_{\rm e}=0.2$, bremsstrahlung
radiation dominates the hard X-ray emission.

\begin{figure}
\centerline{\psfig{figure=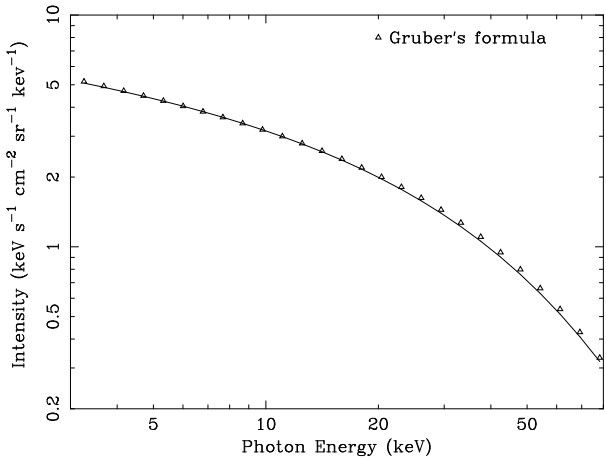,width=0.5\textwidth,angle=270}}
\caption{The X-ray background from sources with two-temperature torus disks at their centres (solid line) compared to Gruber's (1992) empirical formula (Eqn. 19) from 3 to 60 \keV. We take into account the contribution of about 20 per cent from a power-law component of energy index $\sim 0.7$ mostly due to Seyfert galaxies. The temperature of the electrons $\theta_{\rm e}= 0.20$, the  evolution parameter $p=2.6$ and $z_{\rm max}=2.0$. This is the best fit model.}  
\end{figure}

\section{The X-ray background model}

Applying the results reached above we now discuss the emission
spectrum of hot tori as a model for the XRB. The computed value of
$T_{\rm e} \sim 10^9 K$ is nearly independent of the radius $r, m, \dot{m},
\alpha, f$ throughout the two-temperature domain, implying that such
tori would have the same intrinsic spectrum and that thermal
bremsstrahlung would dominate the hard X-ray emission.  We have
calculated the spectral emissivity using the expressions for 
the electron-electron bremsstrahlung and electron-ion bremsstrahlung
computed by Stepney and Guilbert (1983). For the evaluation of the
electron-ion bremsstrahlung, the cross section in the Born approximation
for a relativistic electron in a Coulomb field is given by Gould
(1980). For electron-electron bremsstrahlung process, the
cross-section is much more complicated and we have used the numerically
integrated expression given in Stepney and Guilbert (1983).
\par
 In order to obtain a model for the XRB, we consider the contribution af all sources distributed over a certain redshift range. These X-ray sources are not  resolved as individual sources but they  contribute to the XRB intensity. Here we describe the XRB as a superposition of sources from the local universe ($z=0$) to the early universe based on the framework of the Robertson-Walker metric. We therefore expect the mean luminosity, the number density and the spectra of these sources to change with cosmic time. The total sum of X-rays from the whole population contributes to the XRB we observe locally. The intensity received from objects with local spectral emissivity $j_0$ (comoving $j(z) = j_0(1+z)^{ p}$) distributed up to redshift $z_{\rm max}$ is therefore given by
\begin{eqnarray}
I_{\rm B}(E)(< z_{\rm max})&=&\frac{c}{4\pi H_0}\times\nonumber \\
\\
&&\hspace{-0.5in}\int_{0}^{z_{\rm max}}\frac{(1+z)^{p-2}}{(1+2q_0z)^{1/2}}j_{\rm 0B}(E(1+z))dz,\nonumber
\end{eqnarray}   
where  $q_0$ is the deceleration parameter, $H_0$ the Hubble constant (we use $q_0=0.5$ and $H_0=50 \kmps$) and $j_{0{\rm B}}$ is the total bremsstrahlung  emissivity at the present epoch. $I_B$ represents our computed XRB intensity, which we compare  with the empirical fit to the XRB spectrum from $3$ to $60 \keV$ obtained by Gruber (1992)
\begin{eqnarray}
I_{\rm XRB}& = &7.877E^{-0.29}{\rm exp}(-E/41.13\keV) \nonumber \\
&&\hspace{0.8in}\keV \ps\psr\pcmsq\keV^{-1}.
\end{eqnarray}

We plot Gruber's formula and our model of the XRB in Fig.~4. The fit is obtained by adding a power law component of energy index $\sim 0.7$ to our model. This takes into account a fraction ($\sim 20$ per cent) contributed by AGN (mostly Seyfert galaxies) to the XRB in this energy band (the addition of a such component is not essential and good fits can be obtained by considering the model sources alone). 

We
compare the model to the fit of the XRB by Gruber (1992) in terms of
the fractional errors $\varepsilon =  
 \mid I_B - I_{\rm XRB}\mid/I_{\rm XRB}$
after normalizing at different photon energies. Numerically we find
that the upper limit of the integral, $z_{\rm max} \approx 2$, and
$p\approx 2.6$ give the best fit to the XRB. The spectrum
is not very sensitive to the value of the evolution parameter
$p$, leading to residuals that are always $\approxlt1$ per cent for $2.5\leq
p\leq 3.0$ in the 3-60 \keV ~range. For $p=0$ the  residuals increase
suggesting that some evolution is required for such sources.
\par
We further examine these results by plotting contour levels for the maximum
fractional errors associated with our fit for a given normalization
(which directly corresponds to a given model volume emissivity, $j_{0{\rm m}}$)
and a given $p$ (Fig.~5). We assume that acceptable fits are defined by
the condition max$(\varepsilon) \approxlt 5$ per cent; this condition constrains
the value of the evolution parameter to be between 2.5 - 2.7.
The  model volume emissivity of the source population is then described by
\begin{equation}
j(z)=j_0(1+z)^{2.6\pm0.1}.
\end{equation}
 We find that the
values of $z_{\rm max}$ are plausibly anticorrelated with $p$, low
$z_{\rm max}$ implying high values for spectral evolution parameters for
accectable fits. Relatively high evolution (due to both $z_{\rm max}$
 $\approx 2$ and $p$) seems to fit the peak of the XRB quite well
as it best constrains the model parameters. The precise AGN contribution 
is not important provided it is less than $\sim 30$ per cent at 2 \keV.

\begin{figure}
\centerline{\psfig{figure=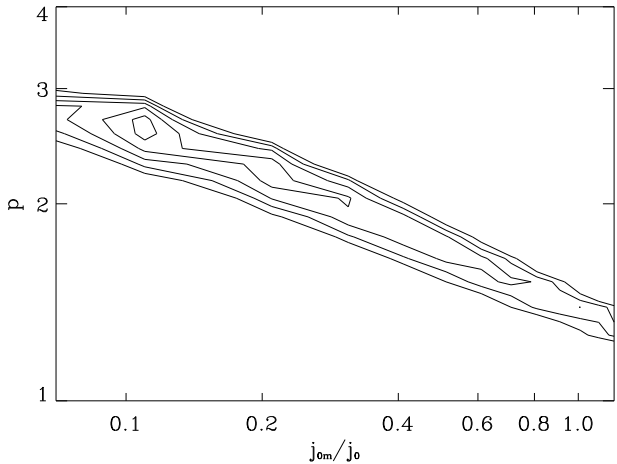,width=0.5\textwidth}}
\caption{ Contour levels for the maximum fractional error, max$(\epsilon)$, calculated for each fit to the XRB obtained by varying $p$, the evolution parameter and  $j_{0{\rm m}}$ is the local emissivity. $j_0$ is the average
 local value of volume emissivity determined by Carrera \etal
 (1995).  The plot constrains the value
 of the evolution parameter and of total contribution to the local
 volume emissivity, i.e. the normalization, to a very well defined
 region of parameter space; implying strong evolution of the sources ($ 2.5\le p\le 2.7$)
 and a contribution of 10 per cent to the local volume emissivity.}
\end{figure}

\subsection{The volume emissivity}

The source population which reproduces the XRB must also be consistent with the local values of X-ray volume emissivity. 
\par
  In recent years it has become apparent that X-ray emission from
  galaxies can make significant contribution to the XRB. Direct
  studies of galaxy contribution using two-point cross-correlation
  between the hard $(2-10 \keV)$ XRB and various optical and infrared
  galaxy catalogues have been carried out (Lahav \etal 1993; Miyajii
  \etal 1994; Carrera \etal 1995) and have provided values for the
  local volume emissivity of the order of $j_0 \sim (3.5 -5.7) \times
  10^{38}h_{50} \ergpspmp $ ($h_{50}=H_0/50\kmpspMpc$). Extrapolation of this
  result back to $z\sim 1-4 $ has lead to the conclusion that
  $\approxlt 10-30$ per cent of the XRB could be produced by a
  population of galaxies if evolution is not included; some moderate
  positive evolution would lead to higher values. 
\par
  We can
  estimate a value for the local X-ray volume emissivity by
  integrating the model spectral emissivity over photon energy. In
  this way we obtain $j_{0{\rm m}} = <n_0L_0> \approx 0.6 \times10^{38}h_{50}
  \ergpspmp$ in the $2-10 \keV$ range implying that our model
  population would be contributing  a fraction of $\approx
  0.10-0.15$ to the local volume emissivity $j_0$ (Fig.~5). 
  If we consider that field galaxies have a spatial density $<n_0>\sim 10^{-3}\pmpccu$, we can determine  the mean X-luminosity of the nearby galaxies cross-correlated with the XRB to be of the order of $<L_0>\approx 6\times10^{40}\ergps$ (in the 2-10\keV ~band), about an order of magnitude higher than that of a typical spiral at the present epoch (Fabbiano 1989). Much of the X-ray luminosity of local spirals is however due to X-ray binaries. This suggests that luminous ion-supported tori are rare at the current epoch. 
\par
Other determinations of  volume emissivity have come from the
 studies of cross-correlation of faint galaxies catalogues with
 unidentified deep ROSAT X-ray sources (Roche \etal 1995; Almaini 1996; Treyer \& Lahav 1996)
 as it has emerged recently that faint galaxies could be
 important contributors to the XRB (see Introduction).  The estimated value of the 
 volume emissivity for a galaxy sample at a median redshift of
 $\overline{z} =0.27$, is $j_0 \sim(2.3)\times 10^{38}h_{50} \ergpspmp$ in
 the $0.5$ to $2 \keV$ range (Almaini 1996, Treyer \& Lahav 1996) . The value we  deduce from our model
 for this energy band is $j_{0{\rm m}} = <n_0L_0> \approx
 1.2\times10^{38}h_{50} \ergpspmp$ where we have considered a sample of
 model sources at the same redshift.  Almaini (1996) has 
 constrained the evolution of the X-ray emissivity with redshift,
 obtaining evidence that X-ray emission from such galaxies evolves as
 strongly as AGN. These results are in good  agreement with our model
 ($p\approx 2.6$).  

 A property of these X-ray emitting galaxies is
 that the X-ray luminosity, $L_{\rm x}\approx 10^{42}\ergps$ (Boyle \etal 1995; Griffiths \etal 1995; Almaini 1996), is typically  two order of magnitude
 more than that of  galaxies observed locally. If our model sources had
 this mean luminosity their number density would be of the order of
 $<n_0>\approx 10^{-4} - 10^{-5}\pmpccu$. These values of $<n_0>$ constrain 
the spatial density of the model population to be very close to that of other AGN at $\approx 1/100$ the  spatial density of local field galaxies.
This determination is strengthened by the fact that the sample analyzed by  Roche \etal (1995) and  Almaini \etal (1996) goes deeper in redshift and possibly decreases the errors associated with the uncertain evolutionary properties of the model galaxies.
 Both the lower luminosity objects observed locally and
 the  higher luminosity population at $\bar{z}=0.3$ are in agreement with our
 ion-torus scenario in which  evolution is important, the
 accretion rate is typically low, $\dot{M} \le 1$ per cent of  $\dot{M}_{\rm Edd}$ and
 the black-hole mass is of the order of $M=10^{7}-10^{9}M_{\odot}$
 (the hard-spectrum is essentially independent of black-hole mass).
\subsection{Radio emission}
The bulk of the emission from the ion-supported tori, occurs at radio and submillimeter band near the self-absorption frequency $\nu_{\rm t}\approx 6\times 10^{11}$ Hz (see Fig.~3). Directly from Fig.~3 we determine a mean radio flux density for the ion-supported tori model at $\approx 10^9$ Hz and compare it with the results of deep VLA images of radio sources. Hamilton \& Helfand (1993) have found some corrrelation of faint radio sources at $20\cm$ with XRB fluctuations  and associated them with the unresolved XRB.
We find that our model predicts a flux density of the order of $\approx 0.3$ mJy at $10^9$ Hz (near $20\cm$) for a source at $z=0.3$. The derived  value of the flux density is consistent with the lower end of the  range of flux densities found by  Hamilton \& Helfand (1993).
\par
Radio surveys at high frequencies (closer to the peak of the self-absorption frequency) are potentially best for selecting ion-supported tori. They could  confirm our model if the radio sources correlate with faint, non-AGN, X-ray sources and have rising radio spectra.

\section{Summary and discussion}
The diffuse XRB in the few-100\keV  ~band can be well explained
 as due to a population of apparently normal galaxies undergoing two-temperature advection-dominated accretion via an ion-supported torus. 
\par
Here, we have considered a physical model for the description of the high energy emission and the cosmological evolution of such sources. We find that a population of sources with X-ray emission  due to an optically-thin thermal plasma emitting bremsstrahlung at a temperature $\theta_{\rm e}\approx0.2$  satisfies the background constraints. We show that the temperature requirement is plausibly achieved in ion-supported tori.
\par
The spectral emissivity of such sources is most likely to be associated with (some) spectral evolution and the acceptable evolution parameters are roughly consistent with those of AGN. The predicted population of sources also satisfies the constraints of the local volume emissivity. In particular we have found that the predicted emissivity  contributes a small fraction to the local volume emissivity in the 2-10\keV ~band. For the deeper ROSAT sample, the  spectral emissivity extrapolated in the 0.5-2\keV
 ~more closely  resembles the value obtained with our model, suggesting that advection-dominated disks might be related to the activity of X-ray luminous, narrow-line galaxies. 

We  note that the low radiative efficiency in the advection-dominated scenario may indicate that most galaxies might harbour a massive black hole as a remnant of earlier more active quasar epoch (see also Fabian \& Rees 1995). In particular, if we suppose that QSOs contribute  about 20 per cent of the XRB at 1 \keV ~and  have power-law spectrum  of energy index $\sim 1$, we can  calculate a ratio of the total emission due to QSOs to that of ADD sources of $E_{\rm QSO}/E_{\rm ADD}\approx 1/4$. If we now assume  that the  radiative efficiency $\eta$ of QSOs and ADD tori to be about 10 per cent and $0.1$ per cent  respectively, we derive an upper limit for the ratio of the mass accretion rates, i.e.,
\[ \frac{\dot{M}_{\rm QSO}}{\dot{M}_{\rm ADD}}\approxlt \frac{1}{4}\left(\frac{\eta_{\rm ADD}}{0.1}\right) \approxlt \frac{1}{400}.\]
This leads directly to the conclusion that  the mean  mass of `dead' black holes might be up to a factor of 100 greater than that estimated previously  from measurements of the luminosities. 
Moreover, given that most of the XRB could be due to a population of ion-supported tori, we can find limits for the total mass density, $\rho$, of black-holes distributed in local galaxies (Soltan 1982 or Fabian \& Canizares 1988). The energy density due to accretion is given by $\varepsilon=\rho\eta c^{2}$  which is equivalent to the the total energy emitted by all of the model sources in a unit comoving volume; i.e. the model XRB emission. If we take the spatial density of host galaxies to be $\approx 10^{-3}\pmpccu$,  we find that most  normal galaxies should contain a central black hole with a mass $\approx 10^{7} M_{\odot}$. Lower space densities lead of course to higher mean masses.
\section{Acknowledgements}
TDM is grateful to members of the IoA X-ray group and in particular Chris Reynolds and Stefano Ettori for many useful conversations.
TDM thanks Trinity College and PPARC for support. ACF thanks the Royal Society
for support.

\end{document}